\documentclass[doublecol]{epl2} 

\usepackage{graphics}
\usepackage{bm}
\usepackage{amsmath}
\usepackage{comment}

\def\Tr#1{{\ensuremath{\text{Tr}{(#1)}}}}
\def\vec#1{{\ensuremath{\bm{#1}}}}
\def\n{\ensuremath{\bm{n} }}

\def\tensor#1{{\ensuremath{\bm{#1}}}}
\def\half{{\textstyle \frac{1}{2}}}

\def\threehalf{{\textstyle \frac{3}{2}}}

\title{Giant deformations and soft-inflation in LCE balloons}
\shorttitle{Title} 

\author{Andrea Giudici\inst{1} \and John S. Biggins \inst{1}}
\shortauthor{A. Giudici and J.S. Biggins}

\institute{Department of Engineering, University of Cambridge, Trumpington St., Cambridge CB21PZ, U.K.                   
}

\abstract{ We propose that ballooning can be controlled, enriched and amplified by using rubbery networks of aligned molecular rods known as liquid crystal elastomers (LCEs). Firstly, LCEs are promising artificial muscles, showing large spontaneous deformations in response to heat and light. In LCE balloons,  spontaneous deformations can trigger classic ballooning, either as phase-separation (at constant volume) or a volume jump (at constant pressure), resulting in greatly magnified actuation strains. Secondly, even at constant temperature, LCEs have unusual mechanics augmented by soft-modes of deformation in which the nematic director rotates within the elastomer. These soft modes enrich the mechanics of LCE balloons, which can also ``balloon'' between rotated and unrotated states, either during the classic instability, or as a separate pre-cursor, leading to successive instabilities during inflation.  }

\begin{document}
\maketitle

\section{Introduction}
Cylindrical balloons, commonly encountered at parties, have $N$ shaped pressure-volume curves, and the negative gradient generates classic ballooning instabilities during inflation \cite{mallock1891ii}. Under pressure control, the balloon jumps in volume at the pressure maximum, to a substantially larger (ballooned) state. Under volume control, the cylinder instead phase-separates into ballooned and un-ballooned portions \cite{ChaterHutchinson, gent2005elastic, meng2014}. Here, we show this instability can be controlled, enriched and amplified in balloons made from liquid crystal elastomers (LCEs).

\begin{figure}[t]
\includegraphics[width=\columnwidth]{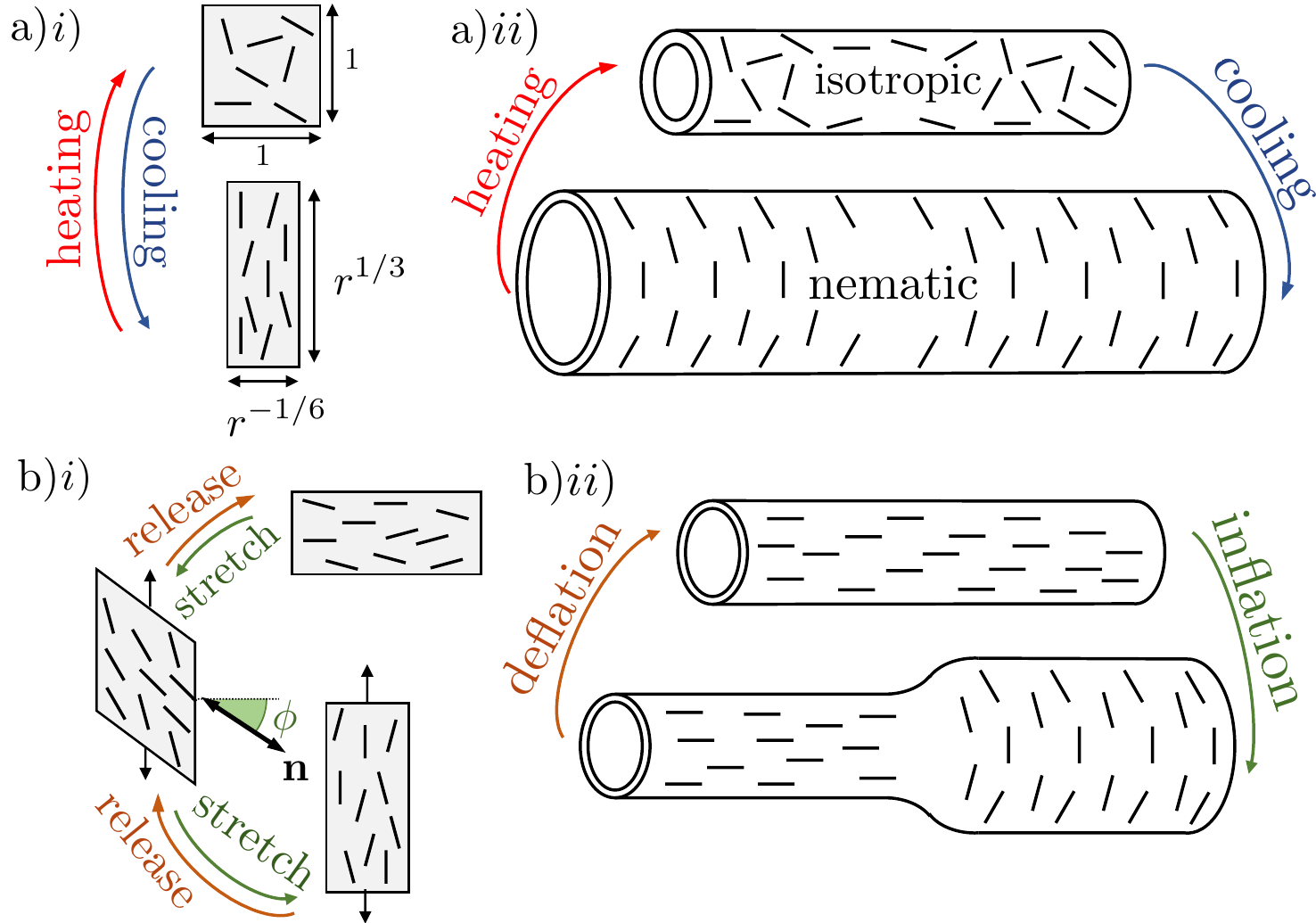}
\caption{Top: LCEs elongate on cooling from isotropic to nematic. In an LCE balloon cooled at constant pressure, this triggers a giant sub-critical volume jump. Bottom: Under stretching an LCE strip can director-rotate, softening the response.  Under inflation, reorientation can become sub-critical, leading to phase separation between rotated and un-rotated segments.}\label{fig.intro1}
\end{figure}

LCEs \cite{warner2007liquid} are rubbery networks of rod-shaped mesogens. Like conventional liquid crystals\cite{de1993physics}, the rods adopt an isotropic orientation distribution when hot, but align below a critical temperature to form a nematic phase. In elastomers, alignment causes a dramatic reversible elongation along the (unit) director $\n$, (Fig.\ \ref{fig.intro1} (a)i), making LCEs soft actuators \cite{kupfer1991nematic,de1997artificial}. LCE bubbles/balloons have  been fabricated \cite{LCEBalloons} but their instabilities remain unexplored.  We show that LCE thermal actuation can trigger the ballooning instability (Fig.\ \ref{fig.intro1}(a)ii), transforming LCEs into sub-critical  actuators with greatly amplified  strain.

LCEs also have very unusual mechanics in the nematic state, stemming from director rotation within the elastomer. In a hot perfectly-isotropic LCE, any director could be chosen on cooling, leading to a degenerate set of ground-states connected by perfectly soft Goldstone-deformations \cite{golubovic1989nonlinear, olmsted1994rotational}. For example, stretch perpendicular to $\n$ (Fig.\ \ref{fig.intro1}(b)) can be accommodated at zero stress entirely by director rotation \cite{finkelmann1997critical, desimone2002macroscopic}. We also consider the inflation of an aligned LCE balloon, and show director rotation is induced towards the dominating azimuthal stress (due to the cylindrical shape) leading to ballooning between fully-rotated and unrotated states (Fig.\ \ref{fig.intro1}(b)). This  may entirely precede the classical ballooning, leading to a balloon with consecutive instabilities during inflation.   

\section{Classical Ballooning} 
\begin{figure}[t]
\includegraphics[width=\columnwidth]{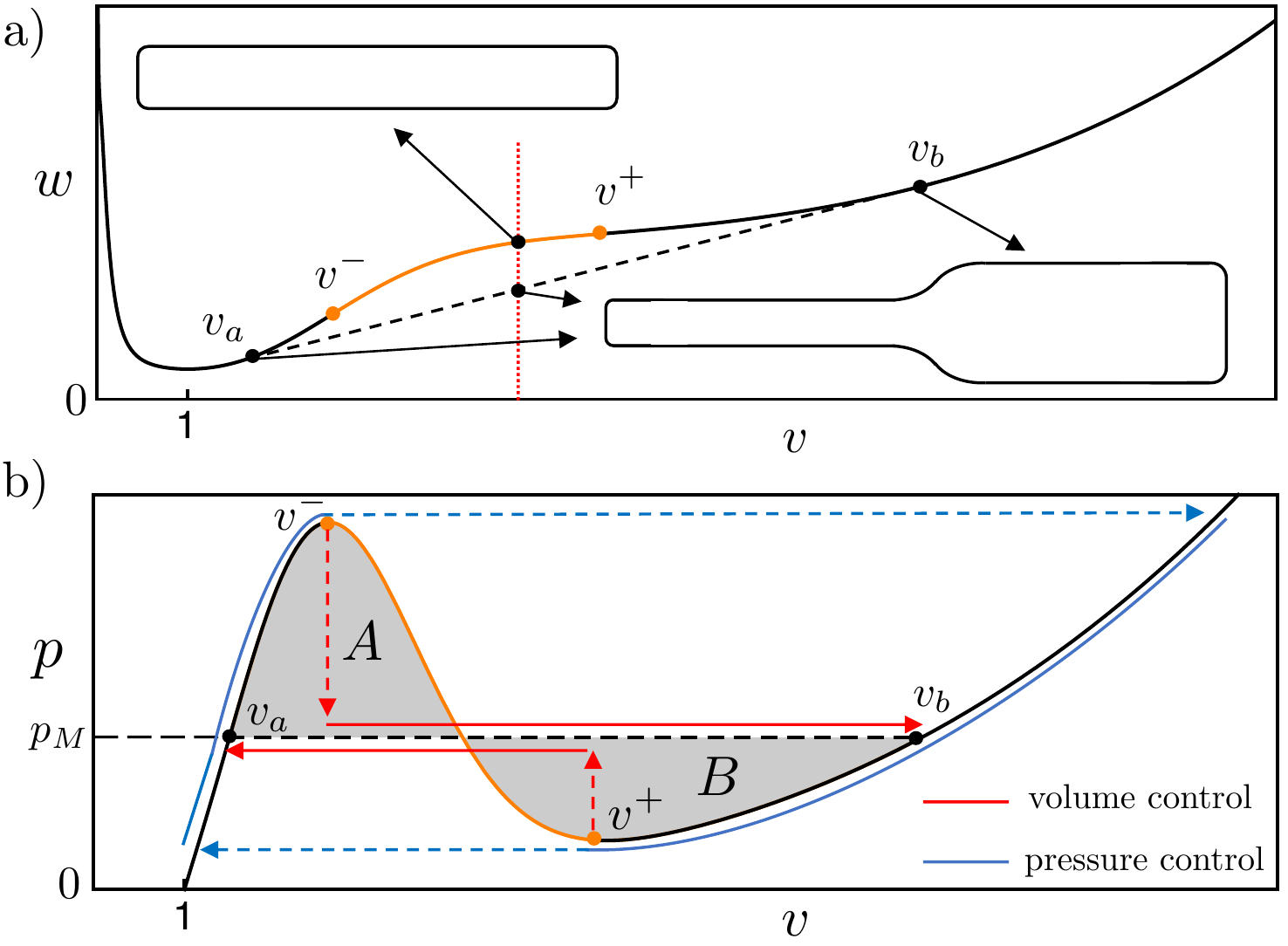}
\caption{ (a) Energy and (b) pressure of a cylindrical balloon, as a function of volume strain $v$. Concavity (orange) in the energy generates an $N$ shaped pressure. Hysteresis loops of volume and pressure control are indicated in red and blue.}
\label{fig. img4}
\end{figure}

We first consider a cylindrical rubber balloon, inflated to a desired volume strain $v=V/V_0$. The balloon instability can be traced to the  elastic energy $w(v)$. As shown in Fig.\ \ref{fig. img4}a), if $w(v)$ has a concave region, $v^-<v<v^+$, and the balloon is inflated into this region, it is advantageous to phase separate (at fixed enclosed volume) into length-fractions at $v_{a}$ and $v_{b}$, as the connecting chord lies below $w(v)$. Optimal separation is achieved by the common tangent construction:
$$w'(v_a)=w'(v_b),\,\,\,\,\,\,\,\,w(v_b)=w(v_a)+w'(v_a)(v_b-v_a).$$
As shown in Fig.\ \ref{fig. img4}b), concavity in $w(v)$ endows the pressure curve, $p\equiv \partial w/\partial V =(1/V_0)w'(v)$, with an unstable negative-gradient, leading to the characteristic $N$ shape. Upon passing $v^{-}$, the balloon phase separates and drops to the Maxwell coexistence-pressure $p_M=p(v_a)=p(v_b)$, which can be found via common-tangents, or the equal area rule for $A$ and $B$. Further inflation is accommodated by enlarging the length-fraction of $v_a$ \cite{ChaterHutchinson}, moving along the energy chord at $p_M$. Phase separation ends at $v_b$, while in deflation it starts at $v^{+}$ and ends at $v_a$. 

However, if the balloon is inflated using a pressure-controlled pump, the form of the instability is quite different. The balloon will dilate homogeneously to $v^{-}$, and then jump  to a fully ballooned state at the same pressure. In deflation,  un-ballooning occurs with a jump from $v^{+}$. 
 
Convexity in $w(v)$ stems from the geometry of large strains. Consider inflating a capped cylindrical membrane so that its length increases $L \to \lambda L$, its radius increases, $R\to\eta R$, and, since the rubber  is incompressible, membrane thickness decreases, $H\to H/(\eta \lambda)$. In cylindrically-oriented locally-Cartesian coordinates $(\hat{\vec{z}}, \hat{\vec{\theta}}, \hat{\vec{\rho}})$, the membrane's deformation gradient is $\textbf{F}=\mathrm{diag}(\lambda, \eta,1/(\eta \lambda))$, and the enclosed volume increases from $V_0=\pi  R^2  L$ to $V= \pi (\eta R)^2 \lambda L$: a volume strain $v\equiv V/V_0=\eta^2 \lambda$.

The simplest model of rubber elasticity treats its  polymers as infinitely-extensible Gaussian chains \cite{wang1952statistical}, leading to the neo-Hookean energy density, which is the square-sum of the principle stretches, $W(\tensor{F})=\half \mu \Tr{\textbf{F}\cdot \textbf{F}^T}\equiv \half \mu I_1$. A neo-Hookean balloon with shear modululus $\mu$ and (fixed) volume of rubber $2 \pi R H L$ thus stores energy
\begin{equation}
w_0={\mu \pi R H L}\left( \eta ^{-2} \lambda ^{-2}+\eta ^2+\lambda ^2 \right).
\label{eqn:nh1}
\end{equation}
However, during inflation, we do not control $\eta$ and $\lambda$, but  the inflationary volume strain $v$. We thus substitute $\lambda=v/\eta^2$ and  set $\eta$ to its minimising value ($\eta_{min}= (2 v^4/(v^2+1))^{1/6}$) to obtain energy and pressure as a functions of $v$:
\begin{align}
w_0(v)&=3\mu \pi R H L\left(\half v+\half v^{-1}\right)^{2/3}\\
p_0(v)&=(\mu H/R)\left(\half v + \half v^{-1}\right)^{-1/3}\left(1-v^{-2}\right).
\label{eqn:neohook}
\end{align}
We now see the cause of ballooning: $w_0(v)$ is concave beyond $v_0^-\equiv \sqrt{4+\sqrt{21}}$, \cite{giudici2020ballooning}, giving a pressure maximum $p_0^-=0.749...\mu H/R$. Ballooning is indeed geometric: at large $v$, membrane strains scale as $v^{1/3}$, giving a concave energy $\propto v^{2/3}$. This simple model captures ballooning's onset, but $w_0(v)$ does not regain convexity at high $v$, so $v_b$ is divergent.  To correct this, we must account for finite chain extensibility, which impose a finite extensibility on the rubber. This is best done using the Gent energy \cite{gent1996new},
 \begin{equation}
W_G(I_1)=-J_m \log \left(1-(I_1-3)/J_m\right),
 \label{eqn:genten}
 \end{equation}
where $J_m$ is a phenomenological limiting value of the strain measure $I_1-3$. Up to an additive constant, Gent matches neo-Hookean for small strains, ($I_1-3 \ll J_m$) but it always regains convexity \cite{gent2005elastic,meng2014}, suggesting that the amplitude of $v_b$ is limited by the finite chains. Unfortunately with Gent, analytic results for $v^{\pm}$, $p_M$, $v_a$ and $v_b$ are not available. However, in a typical rubber balloon, $J_m \sim 80$ is very large, so the onset of instability at $v^{-}$ is well described by neo-Hookean. In most of what follows, we will thus restrict attention to neo-Hookeans, with the implicit assumption that convexity is regained at high $v$. However, since both energies depend on $\tensor{F}$ only via $I_1$, the same (local) deformations always minimise both energies (e.g.\ same $\eta_{min}$), so one may compute Gent energy/pressure curves by substituting neo-Hookean fields into $W_G$.

\section{Nematic-Isotropic Ballooning}
We now consider an LCE balloon, prepared with an azimuthal director, as shown in Fig.\ \ref{fig.intro1}a. This monodomain balloon could be prepared Finkelman-style \cite{kupfer1991nematic} by cross-linking under inflation\cite{LCEBalloons} (so the dominating azimuthal hoop stress imprints azimuthal alignment) or by rolling a planar monodomain. 

\begin{figure}[t]
\includegraphics[width=\columnwidth]{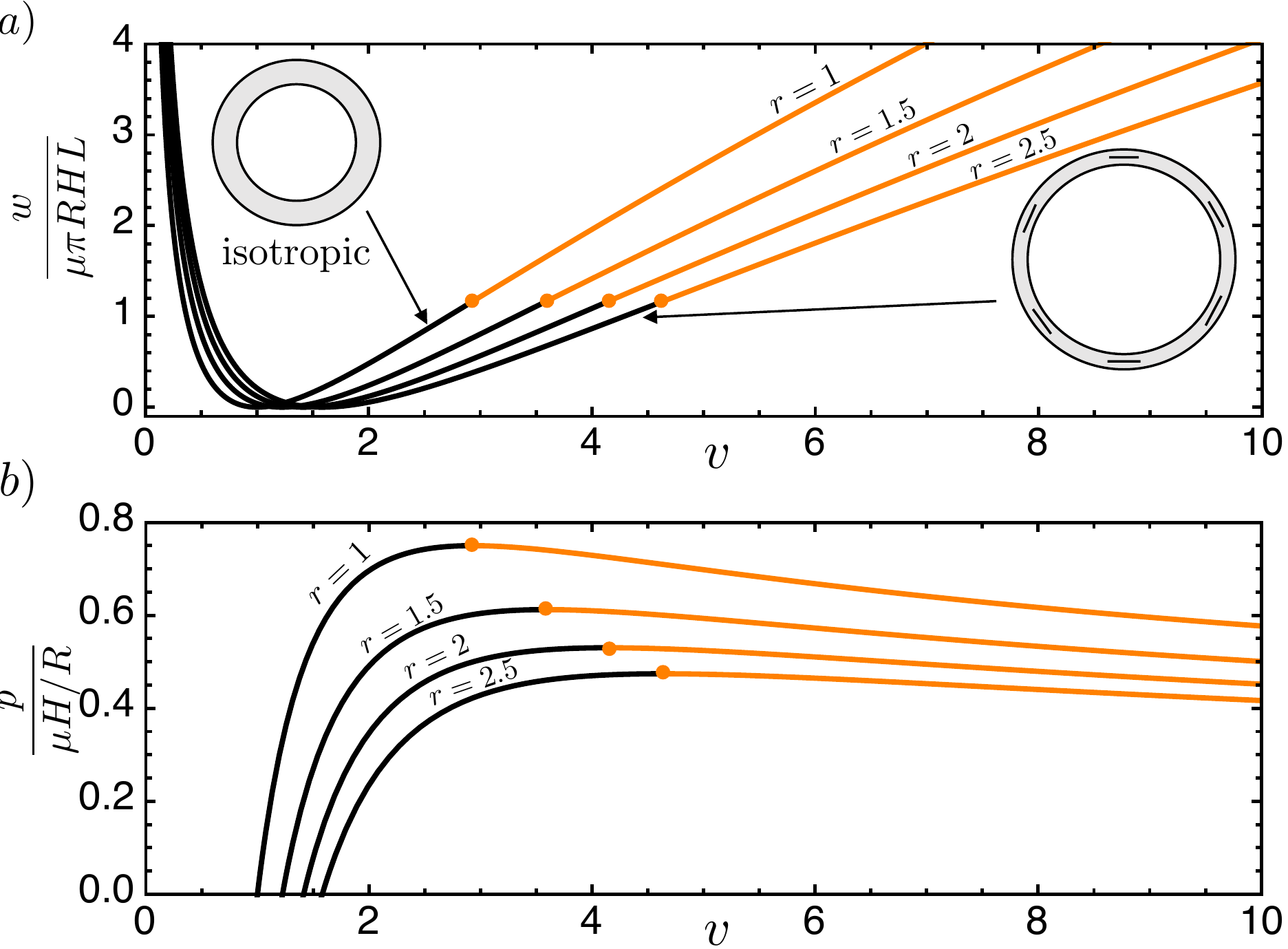}
\caption{Neo-Hookean energy of a cylindrical balloon as a function of volume strain $v$ for different values of $r$. The concave region in the energy, corresponding to a negative gradient in the pressure, is highlighted in orange.}
\label{fig. nemiso}
\end{figure}

For simplicity, we take the hot isotropic state of the balloon as the elastic reference state, with length $L$, radius $R$ and thickness $H$. While isotropic, the LCE behaves just like a conventional rubber, with energy $w_0(v)$ and pressure $p_0(v)$ as above. However, on cooling, the rods align and bias the polymer configurations into prolate forms extended along $\n=\vec{\hat{\theta}}$. Microscopically, these chains are described by a prolate ``step-length-tensor'' which describes the bias of their random walks \cite{warner2007liquid}:
\begin{equation}
\tensor{l}= r^{-1/3}\left( \tensor{\delta}+(r -1) \vec{n}\vec{n}\right).
\label{eqn:l}
\end{equation}
The phenomenological chain anisotropy parameter $r$ subsumes the degree of nematic order and nematic-polymer coupling, while the pre-factor $r^{-1/3}$ ensures $\mathrm{Det}{\left(\tensor{l}\right)}=1$.  A natural generalisation of the neo-Hookean  stat-mech argument then leads to the LCE energy density \cite{warner2007liquid}
\begin{equation}
W(F)=\half \mu \Tr{\tensor{F}^T.\tensor{l}^{-1}.\tensor{F}}.\label{spon_dist_en}
\end{equation}
When the LCE is isotropic, we have $r=1$ and $\tensor{l}=\tensor{\delta}$: standard neo-Hookean. However, on cooling, $r$ grows above one. As illustrated in Fig.\ \ref{fig.intro1}(a)i, the energy is then minimised by the spontaneous deformation $\tensor{F}_s=\tensor{l}^{1/2}=\mathrm{diag}\left(r^{-1/6},r^{1/3},r^{-1/6}\right)$, i.e.\ an elongation by $r^{1/3}$ along $\mathbf{n}$. In principle $r$ is also a function of stress, but this effect is very modest in LCEs owing to the numerous rods per cross-link. The energy has a  multiplicative-decomposition structure, familiar from elastic models of growing tissues \cite{rodriguez1994stress}. Accordingly, if we substitute  $\tensor{F}=\tensor{F_2}\cdot \tensor{F_s}$ (and fix $\n=\vec{\hat{\theta}}$) we recover the original neo-Hookean energy, but for deformations $\tensor{F_2}$ from the spontaneously distorted state.

If the balloon were cooled at zero pressure, the spontaneous distortion would cause it to adopt a new relaxed shape with dilated radius, $R \to r^{1/3} R$, diminished length $L\to r^{-1/6} L $ and diminished thickness $H \to  r^{-1/6} H$: a spontaneous volume strain $v_s=\sqrt{r}$. As above, we decompose the volume strain as $v=v_2 v_s$, with $v_2$ being the strain from  this relaxed state. On inflation, the energy is thus simply $w_0(v_2)$, which, in terms of $v$, gives:
\begin{equation}
w_s (v)=w_0\left(v/\sqrt{r}\right),\,\,\,\,\,\,p_s(v)=p_0(v/\sqrt{r})/\sqrt{r}.
\label{eqn:nemiso}
\end{equation}
As the balloon cools, the energy minimum moves to $v=v_s=\sqrt{r}$, and the instability threshold similarly moves  to $v^-=v_0^- v_s $ (Fig \ref{fig. nemiso}a). These are geometric consequences of the larger relaxed volume. The pressure curves, Fig \ref{fig. nemiso}b, similarly dilate along the $v$ axis by $\sqrt{r}$, but also diminish in height by $\sqrt{r}$ due to the reduced relaxed thickness.

\begin{figure}[h]
\includegraphics[width=\columnwidth]{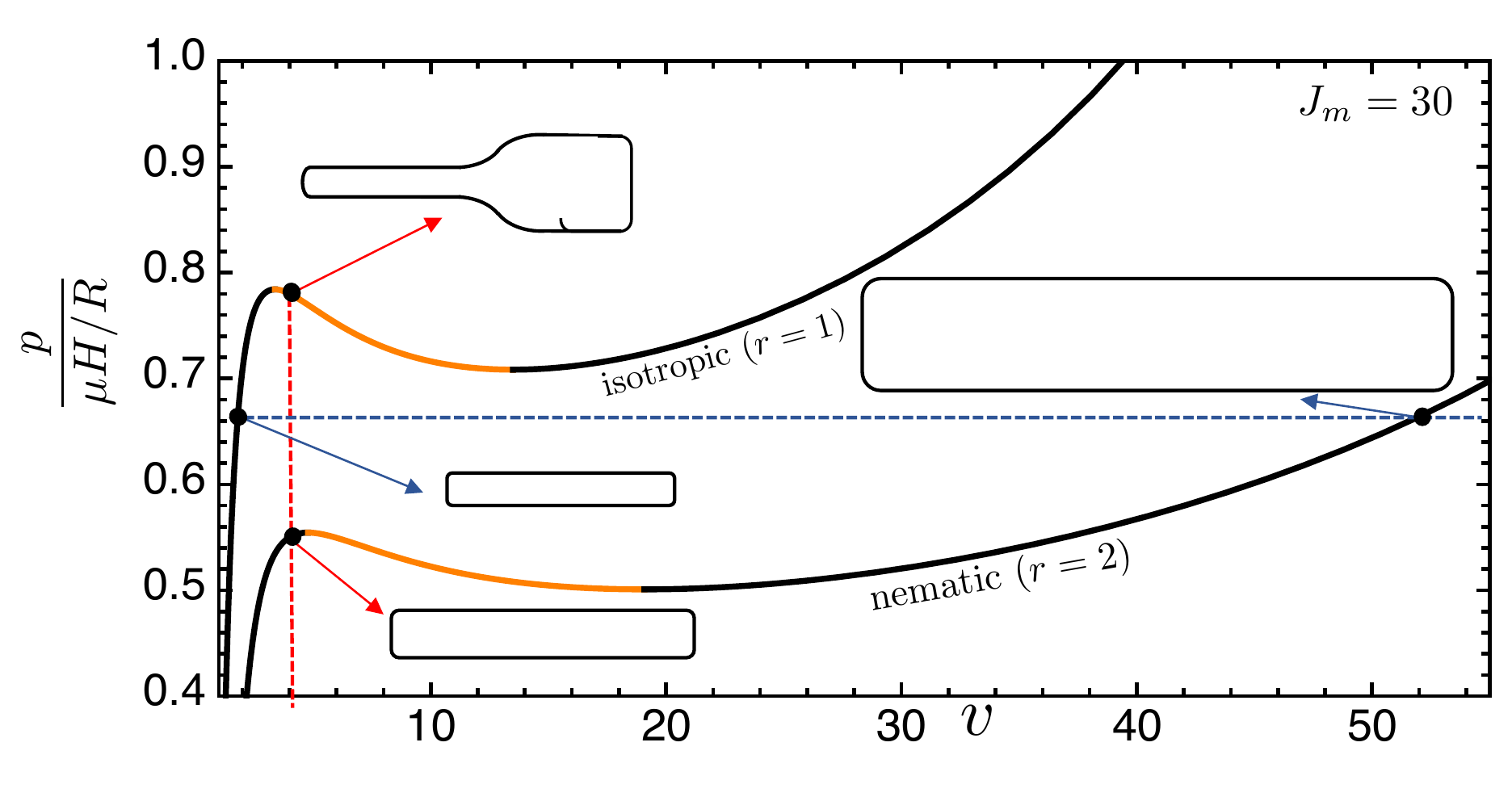}
\caption{Gent pressure curves for the isotropic (hot)  nematic (cold) state with $r=2$. Phase separation is triggered on heating at constant volume (red dashed line) while a giant sub-critical volume jump is triggered on cooling at constant pressure (blue dashed line). }
\label{fig.behaviour}
\end{figure}

These observations lead to two possibilities for temperature controlled ballooning. If we heat at fixed volume the leftward shift of the instability point will cause the balloon to suddenly phase-separate when $v=v_0^-\sqrt{r}$. This possibility is shown as the red isochor on Fig.\ \ref{fig.behaviour}, where we now deploy Gent pressure curves predicting finite amplitude. Conversely, if we cool at constant pressure, (blue isobar) the diminishing height of the $p-v$ curve can cause the LCE to jump  to a fully ballooned state when $p=p_0^-/\sqrt{r}$.

In both cases, the LCE actuation is transformed. To create an actuating mono-domain, one must imprint a preferred director through the elastomer, either by stretching during cross-linking  \cite{kupfer1991nematic} or by cross-linking in an aligned liquid nematic state \cite{broer1995creation,ware2015voxelated}. Imprinting renders mono-domain LCEs supercritical, with continuous actuation over a considerable temperature range \cite{warner2007liquid}. However, ballooning  transforms LCEs into sub-critical actuators: the strain in the LCE jumps as the ballooning threshold is passed. Furthermore, the strain in the ballooned state is $v_b \sim J_m^{3/2}$, reflecting the finite extensibility of the polymer chains, and far larger than the LCE's intrinsic actuation strain.  A similar principle has been deployed to amplify voltage driven actuation in dielectric elastomer balloons \cite{rudykh2012snap, li2013giant}. Such deformations may be termed giant, as their limiting value reflects different physics to their origin. 

A simpler option may be to take an LCE balloon crosslinked in the high-$T$ isotropic state without imprinting $\n_0$. The nematic-isotropic transition is now first order, reflecting the non-polar nature of the nematic phase \cite{de1993physics}. On cooling, such LCEs form `isotropic-genesis' polydomains, without macroscopic actuation. However, such samples are ``super-soft'' \cite{urayama2009polydomain, biggins2009supersoft} with even a very slight stress being sufficient to align the director and deliver the monodomain deformation $\sim r^{1/3}$. In a pressurized balloon, the dominating azimuthal stress would guide the director azimuthally on cooling, leading to above actuation pathways, but now triggered by a discontinuous change in $r$ as the first-order phase-transition is traversed.

\section{Soft-mode Ballooning}
Upon cooling a (hypothetical) perfectly-isotropic LCE monodomain, any nematic director $\n_0$ could be chosen, and each possibility produces a different but equivalent spontaneous deformation $\tensor{F_s}=\tensor{l}_0^{1/2}$. This broken-symmetery endows cold LCEs with Goldsone-like soft modes of deformation which map the LCE between different but equivalent ground states, accompanied by rotation of the alignment to a new director $\n$. Indeed, if we substitute $\tensor{F} \to \tensor{F}\tensor{l}_0^{1/2}$ into eqn.\ \ref{spon_dist_en} (so that $\tensor{F}$ is now the deformation of the aligned nematic LCE with $\n_0$) the elastic energy becomes \cite{warner2007liquid}
\begin{equation}
    W_s(\mathbf{F},\n)= \half \mu \Tr{\tensor{l}_0.\tensor{F}^T.\tensor{l}^{-1}.\tensor{F}}\label{eq:nem_energy},
\end{equation}
where the current director $\n$ is now understood as free to rotate within the LCE. This energy is minimised by any distortion in which the director rotates to $\n$ and the elastomer undergoes the Goldstone deformation $\tensor{F}_{soft}=\tensor{l}^{1/2}\tensor{R}\tensor{l}_0^{-1/2}$,  with $\tensor{R}$ a rotation. Most familiarly, stretch perpendicular to $\n_0$ can be accommodated softly (up to $\sqrt{r}$) via reorientation \cite{finkelmann1997critical, warner2007liquid}, as sketched in Fig.\ \ref{fig.intro1}(b)i.

Similarly, upon cooling an isotropic LCE balloon at zero pressure, any direction could be chosen for $\n_0$, and different choices give different spontaneous deformations, with different volume strains. As previously, an azimuthal director produces a spontaneous dilation $v_s=\sqrt{r}$. In contrast, longitudinal $\n_0$ would give $L \to r^{1/3} L$ and $R \to r^{-1/6} R$ with no overall volume strain;  homeotropic (radial) director would give $R \to r^{-1/6} R$ and $L \to r^{-1/6} L$ for an overall contraction $v_s=1/\sqrt{r}$; and oblique director $\n_0$ would interpolate these three limits. If one inflates an initially homeotropic balloon, it will traverse these states via director rotation (at zero pressure) until it reaches the azimuthal state, which encloses the largest volume. We thus  encounter a ``soft'' zero-pressure dilation up to $v=r$ (relative to the homeotropic initial state) followed by conventional inflation of the azimuthal balloon until the classical ballooning threshold is reached at $v=v_0^{-} r$. Similarly,  a longitudinal balloon will inflate softly to $v=\sqrt{r}$, then inflate with $\n$ azimuthal until ballooning at $v=v_0^{-} \sqrt{r}$.

 However, although monodomain LCEs do deform very softly via director rotation, the stress is never quite zero, and they do return to their original configuration on release. The origin of this non-ideal (``semi-soft'') behaviour is the alignment required during monodomain fabrication, which imprints a preferred director orientation and breaks the  degeneracy. This behaviour is captured by a``semi-soft'' addition to the nematic energy that encodes a preference for the current director $\n$ to align with the imprinted director $\n_0$ via a non-ideality parameter $\alpha$:
\begin{equation}
W(\tensor{F},\n)=W_s(\tensor{F},\n) +\half \mu \alpha \text{Tr}\left((\tensor{\delta}-\n_0\n_0).\tensor{F}^T.(\n\n).\textbf{F} \right). \notag
\label{Fnh}
\end{equation}
This addition can be justified microscopically \cite{verwey1997compositional} or phenomenologically \cite{biggins2008semisoft}, and provides an excellent description of director rotation in LCE strips \cite{finkelmann1997critical}. In particular, in the iconic perpendicular stretch experiment (Fig.\ \ref{fig.intro1}(b)i),  director rotation is delayed until a threshold stretch $\lambda_i=((r-1)/(r-1-\alpha r))^{1/3}$, then proceeds continuously with increasing stress/stretch (leading to characteristic stress plateau) before completing at $\lambda_f=\sqrt{r} \lambda_i$ \cite{warner2007liquid}.

In LCE balloons, non-ideality is essential to resolve the $p-v$ curve during rotation, and probe stability. We start by considering a balloon ($R$, $L$, $H$) with homeotropic initial director, $\n_0=\vec{\hat{\rho}}$, being inflated by $v$. Prior to rotation, ($\n=\n_0$), the semi-soft energy reproduces the neo-Hookean energy and pressure, $w_0(v)$ and $p_0(v)$, with a minimum at $v=1$ and classical balloon instability at $v^-_0$. However, past threshold, we expect director rotation towards dominating azimuthal stress, $\n_{\phi}=\sin(\phi)\vec{\hat{\theta}}+\cos(\phi)\vec{\hat{\rho}}$. As seen in stretching (Fig.\ \ref{fig.intro1}(b)i, \cite{finkelmann1997critical, warner2007liquid}), during rotation, we must allow sympathetic shears, $F_{ \theta \rho} =s$ so the LCE can access soft Goldstone-type deformations:
\begin{align}
\textbf{F}&=\begin{pmatrix} 
\lambda & 0 & 0 \\
0 &  \eta & s \\
0 &  0 &  1/(\lambda \eta)
\end{pmatrix}.
\end{align}
Shear $F_{\rho \theta}$ could also enable soft deformations, but it makes non-circular cross-sections, and (as $\lambda_{zx}$ in stretched strips \cite{warner2007liquid}) is suppressed  (via torques) by  azimuthal stress.

Once rotation is complete ($\n_{\pi/2}=\vec{\hat{\theta}}$) the shear must vanish (on symmetry grounds) and the semi-soft energy reduces to  $w_{\pi/2}=\mu \pi R H L\left(\eta^2/r+\lambda^2+r/(\eta^2 \lambda^2)+\alpha \eta^2\right)$. During this final portion of inflation we can simply substitute $\lambda=v/\eta^2$ and minimise over $\eta$ to obtain the energy (and hence the pressure) as a function of $v$. This again reveals a re-scaled version of a neo-Hookean balloon:
\begin{align}
w_{\pi/2}(v)&=(1+\alpha r )^{1/3}w_0\left(v/v_{\pi/2}\right)\\
p_{\pi/2}(v)&=(1+\alpha r )^{1/3}p_0\left(v/v_{\pi/2}\right)/v_{\pi/2} 
\label{eqn:wpiby2}
\end{align}
where $v_{\pi/2}=r/\sqrt{\alpha r+1}>1$ is the minimum of the azimuthal energy, which, reassuringly, reduces to $v_{\pi/2}=r$ in the ideal case, reflecting the ideal end of soft inflation.

\begin{figure}[h]
\includegraphics[width=\columnwidth]{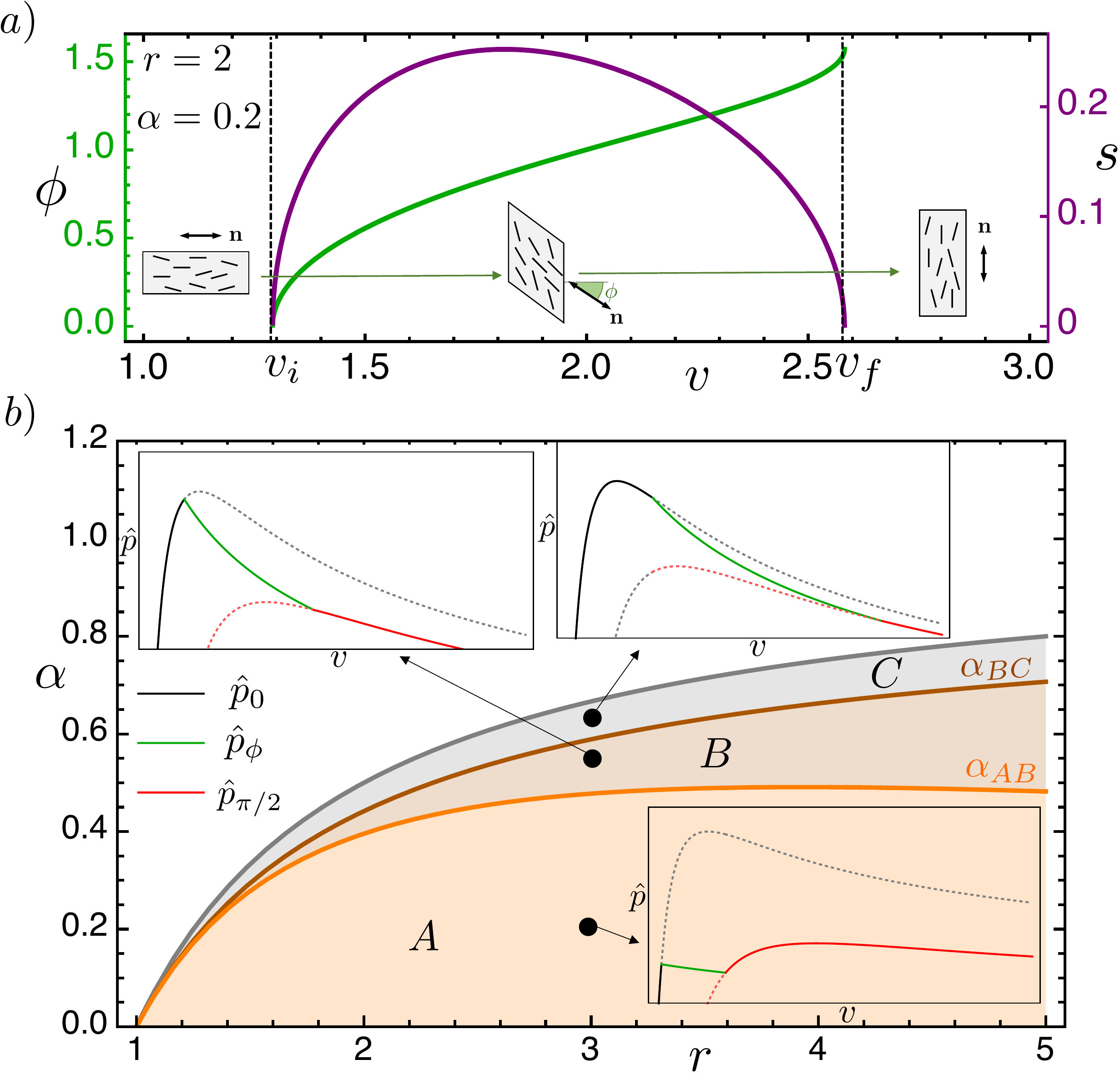}
\caption{ \small a) Director rotation angle $\phi$ (green) and shear $s$ (purple) as a function of $v$. b) Summary of phase separation behaviour as dependent on $\alpha$ and $r$. In section $A$ classical and rotational instabilities happen consecutively. In $B$ there is a single instability, started by director reorientation and saturating via the classic instability. In  $C$ there is a single instability with classical onset and saturation, but decorated by reorientation. Above the limiting line $\alpha=(r-1)/r$  rotation is  suppressed.} 
\label{dirrotation}
\end{figure}

To find the energy and pressure during rotation, we substitute our general $\tensor{F}$ and an oblique director $\n_{\phi}$ into the semi-soft energy. We then further substitute $\lambda=v/\eta^2$ and minimise over $s$, $\phi$ and $\eta$ (detailed algebra in SI) to get
\begin{align*}
    &s=\frac{\eta  (r-1) \sin \phi \cos \phi}{v \left(\sin ^2\phi+r \left(1-\sin ^2\phi\right)\right)}\\
    &\sin^{2}\phi=\frac{r (v-v_i)}{(r-1) v },\,\,\,\,\, \eta=\left(\frac{2v^4}{v^2+2 v v_i^{-1}- v^2 v_i^{-2} }\right)^{1/6}
\end{align*}
where $v_i\equiv\sqrt{\frac{r-1}{r-1-\alpha r}}$. As plotted in Fig.\ref{dirrotation}a, director rotation (and sympathetic shear) proceeds between $v_i$ and $v_f=r\,v_i$, again reflecting the ideal degree of soft-inflation.

The balloon director can rotate towards azimuthal in either sense,  $\pm \phi$, with opposite shear $\pm s$. Stretched LCE strips famously break into ``stripe-domains'' of alternate rotation \cite{finkelmann1997critical}, in order to avoid macroscopic shear at the clamps. Initially homeotropic balloons, only shear  in the thickness direction, so the displacement is small and unconstrained, leaving no imperative for stripes. Different regions may nevertheless rotate in different senses, leading to domain patterns but not changing the mechanics.

Substituting all these fields back into the full semi-soft energy, we can evaluate to energy and pressure:
\begin{align}
w_{\phi}(v)&=3 \mu \pi R H L \left(v_i^{-1}-\half v v_i^{-2}+\half v \right)^{2/3}\\
p_{\phi}(v)&=\frac{\mu H}{R} \left(v_i^{-1}-\half v v_i^{-2}+\half v \right)^{-1/3}\left(1-v_i^{-2}\right).
\end{align}
Reassuringly, both energy and pressure meet their initial (homeotropic) and final (azimuthal) counterparts continuously at $v_i$ and $v_f$ respectively, leading to continuous $p-v$ curves containing three segments, as shown in Fig.\ref{dirrotation}b. Furthermore, we see that $p_{\phi}$ would indeed vanish in the ideal limit $v_i=1$, and should be strikingly small in real LCEs.

Excitingly, the energy during rotation $w_\phi(v)$ is strictly concave, generating a descending pressure curve, and guaranteeing a sub-critical ballooning during rotation. For an almost ideal LCE ($\alpha \ll 1$, region A of Fig.\ref{dirrotation}b) director rotation starts at very small inflation, $v_i\sim 1$, and immediately nucleates a sub-critical ballooning from homeotropic to azimuthal (via phase-separation under volume control, or a volume jump under pressure control). This initial ballooning is entirely LCE in character, and covers a volume strain $v \sim r$. Additional inflation then occurs with azimuthal director, and the classical ballooning instability occurs later, when $v^{-}=v_0^-v_{\pi/2}$ is reached in the azimuthal energy. If $\alpha$ is larger (region B) the onset of rotation is delayed, and the end of rotation joins $p_{\pi/2}$ after the classical ballooning instability, giving a single combined instability nucleated by director rotation, but saturated classically in the azimuthal state by unwinding polymers.  Yet larger $\alpha$ (C) delays rotation beyond the classical ballooning threshold in the homeotropic state, so ballooning starts and finishes classically, but is decorated by director rotation. The lines between these regions can be found by setting $v_f=v_{\pi/2}^-$ and $v_i=v_0^{-}$ respectively, to get
\begin{equation}
\alpha_{AB}=\frac{\left(3+\sqrt{21}\right) (r-1)}{r \left(r+\sqrt{21}+3\right)}, \,\,\,\,\alpha_{BC}=\frac{\sqrt{21}-9}{5}\left(\frac{1-r}{r}\right). \notag
\end{equation}

\subsection{Longitudinal initial alignment}
Finally, we present the results for a balloon with initially longitudinal director, $\n_0=\hat{\mathbf{z}}$. Following the same steps as before (algebra in SI), the initial (unrotated) energy and pressure are still $w_0(v)$ and $p_0(v)$. Similarly, their fully rotated (azimuthal) counterparts are still $w_{\pi/2}$ and $p_{\pi/2}$ from eqn. \eqref{eqn:wpiby2}, but with minimising volume strain $v_{\pi/2}=\sqrt{r/(1+\alpha r)}$, which now recalls the extent of the soft-inflation for an ideal longitudinal balloon, $v=\sqrt{r}$. Between, we expect an oblique rotating director, $\n_{\phi}=\sin{\phi}\hat{\theta}+\cos{\phi}\hat{z}$, which demands inclusion of a sympathetic shear $s=F_{\theta z}$. This shear is in-plane (like stretched strips) leading to macroscopic displacements that cause the balloon to twist along its length. Again their are two senses for rotation and shear, and striping may be encountered to eliminate this macroscopic twist (if the ends of the balloon are constrained) but without changing the $p-v$ mechanics. 

Going through the same minimizations ($s$, $\phi$ and $\eta$) we obtain the energy and pressure during rotation:
\begin{align}
&w_{\phi}(v)=\threehalf \mu \pi R H L \left(2 v^2 \left(1-v_i^{-4}\right)+4 v_i^{-2}+4\right)^{1/3}\\
& p_{\phi}(v)=\frac{\mu H}{R}\frac{2 v\left(1-v_i^{-4}\right)}{\left(2 v^2 \left(1-v_i^{-4}\right)+4 v_i^{-2}+4\right)^{2/3}}
\end{align}
where $v_i=\sqrt{\frac{1-r}{2 \alpha r-r+1}}$ and $v_f=\sqrt{\frac{(1-r) r}{r (\alpha r+\alpha-1)+1}}$ denote the start and end of rotation. In this case, the pressure during rotation rises  up to a maximum  at $v_c=\sqrt{6/\left(1-v_i^{-2}\right)}$ before decreasing, indicating that director rotation can occur continuously up to $v_c$, and sub-critically thereafter.

\begin{figure}[h]
\includegraphics[width=\columnwidth]{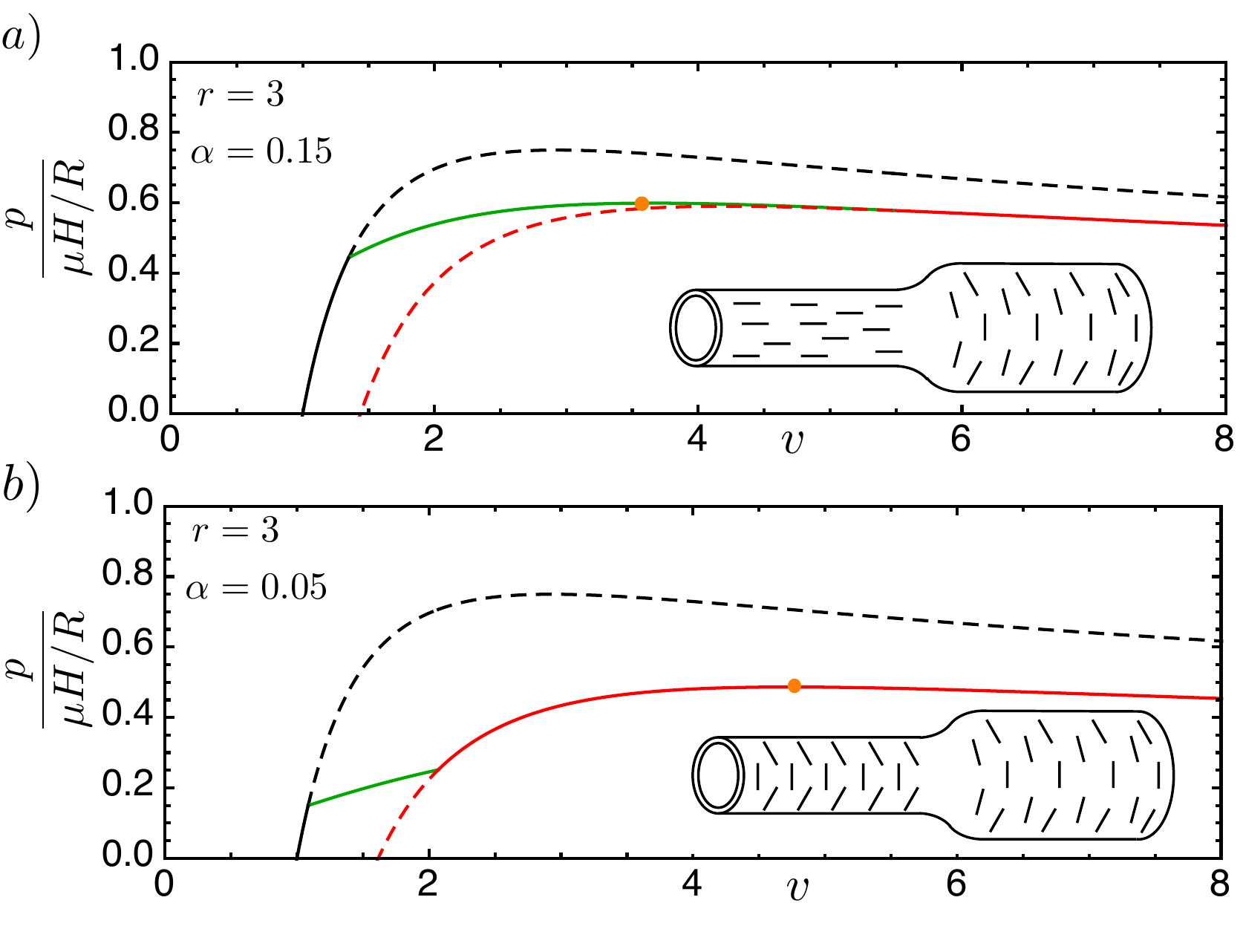}
\caption{Pressure curve for initially longitudinal alignment of the director-field. In the case of large $\alpha$ (a) instability is triggered during director rotation, and phase separation is between regions of different alignment. Small $\alpha$ (b) leads to rotation concluding prior to instability, so phase-separation happens entirely within the azimuthal section of the pressure curve.  }
\label{longitudinal}
\end{figure}

We may again construct 3-section pressure curves to examine the stability of longitudinal systems. In the case of small $\alpha$, Figure \ref{longitudinal}b, we find the energy remains convex throughout rotation, and ballooning only occurs at the classical ballooning threshold in the fully rotated (azimuthal) state. Upon ballooning (under volume control) phase separation will saturate into $v_b$ (with azimuthal director) and $v_a$, which may be azimuthal, rotating or longitudinal depending on the Maxwell pressure, which is in turn determined by $J_m$. However, for sufficiently small $\alpha$, $v_b$ will surely also be azimuthal, so continuouss director rotation will entirely precede azimuthal ballooning. On the other hand, at large $\alpha$, Figure \ref{longitudinal}a, the pressure curve of the rotating section has a maximum at $v_c$, corresponding to an instability. In this case, phase separation will be initiated when the director is partially rotated, and the two phases are guaranteed to have different orientations. The transition between these regimes occurs at: 
\begin{equation}
\alpha=\frac{3 \left(\sqrt{21}-1\right) (r-1)}{r \left(20 r+3 \sqrt{21}-3\right)}
\end{equation}

\section{Discussion} Ballooning is the original large strain elastic instability. However, in recent years, many more examples have been documented: cavitation, wrinkling, buckling, fingering creasing and beading to name but a few. These instabilities are of considerable interest in soft solids, as they can be used to reversibly sculpt shape. Here we have demonstrated two ways in which the ballooning instability can be controlled and enriched by using LCEs. Firstly, we have seen how the thermal strains of an LCE can trigger ballooning. This process leverages LCEs to create balloons that respond to heat and light. However, it also leaverages ballooning to greatly magnifying the LCEs intrinsic actuation, and generate an explosive sub-critical response that could be used for switching, jumping or threshold-sensing. Secondly, we have seen how the soft-modes associated with director-rotation within an LCE can produce entirely new modes of ballooning. We anticipate that LCEs, and other soft actuators, can similarly control and enrich a wide range of classical instabilities, facilitating their deployment in shape-shifting devices. 

\bibliographystyle{eplbib}
\bibliography{references.bib}

\end{document}